\documentclass[reprint,superscriptaddress]{revtex4-1}

\usepackage{array}
\usepackage{graphicx}
\usepackage{verbatim}
\usepackage{color}
\usepackage{multirow}

\newenvironment{methods_summary}{%
    \section*{Methods summary}%
    \setlength{\parskip}{12pt}%
    }{}

\newcommand{\ket}[1]{\left\vert{#1}\right\rangle}

\begin{document}

\title{A high-speed optical link to entangle quantum dots}
\author{Cody Jones}
\email{ncodyjones@gmail.com}
\affiliation{Edward L. Ginzton Laboratory,
         Stanford University,
         Stanford, California 94305-4088, USA}
\author{Kristiaan~De~Greve}
\affiliation{Edward L. Ginzton Laboratory,
         Stanford University,
         Stanford, California 94305-4088, USA}
\author{Yoshihisa Yamamoto}
\affiliation{Edward L. Ginzton Laboratory,
         Stanford University,
         Stanford, California 94305-4088, USA}
\affiliation{National Institute of Informatics,
         Hitotsubashi 2-1-2, Chiyoda-ku,
         Tokyo 101-8403, Japan}

%Currently at Harvard University, Department of Physics, 17 Oxford Street, Cambridge MA 02138, USA\\

\begin{abstract}
A long-distance quantum network for distributing entangled states would support novel information applications, such as unconditionally secure cryptography~\cite{Briegel1998,Gisin2007} and distributed quantum computing~\cite{Cirac1999,Barz2012}.  Realizing such a network requires hardware that can reliably and efficiently establish entanglement over long distances, despite challenges like loss and transmission delay~\cite{Pan1998,Duan2001,Moehring2007,Kimble2008,Hofmann2012,Yin2012,Ritter2012,Jouguet2013,Nauerth2013,Bernien2013}.  We propose a new scheme for distributing entanglement that increases communication rates one to two orders of magnitude over existing protocols.  The method is less sensitive to overall system loss because it transmits many signals within one round-trip-time window and efficiently discriminates lost-photon events.  While the scheme applies to many types of matter qubits, we analyze a specific implementation with optical quantum dots, showing that the method uses practical hardware consistent with recent experiments~\cite{Gao2012,DeGreve2012,Schaibley2013}.
\end{abstract}

\maketitle

Quantum networks could enable transformative technologies in data security and information processing~\cite{Briegel1998,Barz2012}.  However, transmission of quantum signals comes with unique challenges, like the no-cloning theorem, which prohibits the use of amplifiers to overcome signal attenuation~\cite{Gisin2007}.  To address attenuation from absorption or coupling losses, entanglement can be distributed by repeatedly attempting to produce a distributed Bell state, a maximally-entangled pair of qubits that can be used for cryptography~\cite{Briegel1998} or distributed computing~\cite{Barz2012}.  Like many previous proposals~\cite{Briegel1998,Gisin2007,Simon2007,Munro2010}, we present a scheme that generates Bell pairs distributed over a long distance, a crucial procedure in a quantum network.  The advance reported here is an improvement in communication rate through robustness to system loss.  Imagine that any single communication attempt has a success probability $\beta \ll 1$.  Whereas prior schemes had a communication rate proportional to $\beta$, our scheme uses efficient discrimination of lost-photon events and rapid-fire distribution of entangled photons to communicate Bell pairs between two quantum dots at a rate proportional to $\sqrt{\beta}$.  The technique applies equally well to other matter qubits, and several modifications to the optical components are discussed.

We consider a means to generate entanglement between quantum dots, because they are a promising platform for integrated quantum-processing devices~\cite{Imamoglu1999} and because they can be induced to emit a single photon entangled to the quantum dot spin, which is crucial to our proposal.  When a single electron is spatially confined by an InAs quantum dot at 1.5~Kelvin in a magnetic field, the two electron spin states form a qubit, and the presence of two excited trion states yields a four-level energy diagram~\cite{Bayer2002} that facilitates qubit readout and coherent control~\cite{Imamoglu1999,Berezovsky2008,Press2008}.  Importantly, when a pump laser controlled by an electro-optic modulator (EOM) excites the quantum dot into a trion state (Fig.~1), it will emit a single photon by spontaneous emission with polarization entangled to the electron spin left behind~\cite{Gao2012,DeGreve2012,Schaibley2013}.  This photon can be collected with a high-numerical-aperture lens and coupled into an optical channel.

If two charged quantum dots have the same transitions in frequency and polarization, then the dot spins can be entangled by interfering their emitted photons using the Hong-Ou-Mandel effect~\cite{Hong1987,Moehring2007}, as shown in Fig.~1.  This apparatus coherently converts emitted photons to a telecom wavelength around 1550~nm for low-loss fiber transmission, as discussed in Methods.  Without loss of generality, we assume ideal detectors at this point.  The photons interfere in a linear, non-deterministic Bell-state measurement (BSM)~\cite{Zukowski1993,Pan1998} apparatus that lies at the channel midpoint.  When two indistinguishable photons enter the 50:50 beam splitter, they will always bunch together and exit the same port, registering a signal at only one detector.  However, distinguishable photons will exit through either port with 50\% probability, such that a two-detector, ``double click'' event projects the quantum-dot spins into an entangled singlet state $2^{-1/2}(\ket{\uparrow\downarrow}-\ket{\downarrow\uparrow})$, and classical signals inform each side of success.  The process of creating spin-spin entanglement through interference of spin-photon pairs is known as entanglement swapping~\cite{Briegel1998,Pan1998,Duan2001}.  The partial BSM in Fig.~1 succeeds in only 25\% of the instances where two photons arrive. Appendix~\ref{App_BSM} describes how to increase success fraction to 50\%, and Appendix~\ref{App_time_bin} discusses how to perform a partial BSM using time-bin entanglement.  The overall scheme, which we call ``midpoint interference,'' has been experimentally verified in trapped ions~\cite{Moehring2007}, photons~\cite{Pan1998,Yin2012}, atoms~\cite{Ritter2012,Hofmann2012}, and nitrogen-vacancy centers in diamond~\cite{Bernien2013}.

\begin{figure*}
  \centering
  \includegraphics[width=\textwidth]{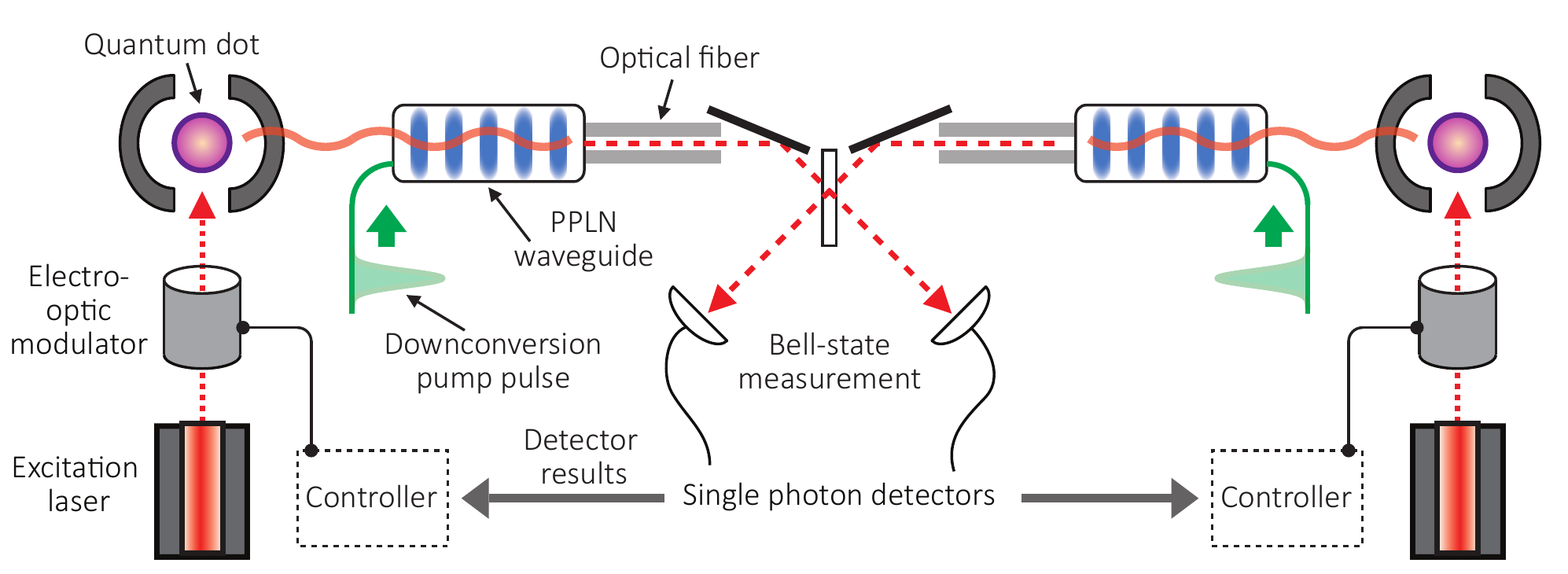}
  \caption{Midpoint-interference scheme.  Description of components moves clockwise from the excitation laser.  The controller can open or close an electro-optic modulator (EOM) to regulate excitation laser pulses.  A pulse excites the quantum dot into producing a spin-photon entangled state through spontaneous emission.  The emitted photon is downconverted to telecom wavelength using a periodically poled lithium niobate (PPLN) waveguide.  The converted photon is coupled into optical fiber, which is perhaps 10--50~km in length.  Photons from both sides of the channel interfere at the Bell-state measurement (BSM) apparatus, located at the midpoint.  A double-click detection event projects the quantum-dot spins into an entangled state.  Classical signals inform each controller of the BSM result, indicating that the quantum dots can attempt entanglement again.}
  \label{Midpoint_interference}
\end{figure*}

When using midpoint interference, each of the quantum dots must hold its spin qubit as idle memory until the detector signals propagate back, and this delay limits performance.  If the length of the channel is 50~km, the transmission delay is $\tau_t = 250$~$\mu$s in optical fiber.  Moreover, the spin qubits are likely to be discarded because of lost photons.  If the total photon loss in dB for the midpoint-interference scheme is $\alpha_1$ (including all inefficiencies in BSM and photon collection) and the probability of not losing photons is $\beta_1 = 10^{-\alpha_1/10}$, then the average rate of entanglement generation using midpoint interference is $G_1 = \beta_1/\tau_t$.  Recent experiments suggest total system loss may be $\alpha_1 =$ 40--60~dB under optimistic assumptions~\cite{Moehring2007,Gao2012,DeGreve2012,Schaibley2013,Bernien2013}, meaning $G_1$ at 50~km is less than one entangled spin pair per second.

Several modifications to the midpoint-interference scheme have been proposed to increase the rate of quantum signal transmission.  Simon et al. proposed using rare-earth-doped crystals with multi-mode storage of photons, allowing many independent transmissions through the same channel~\cite{Simon2007}.  Munro et al. propose an asymmetric design where one side of the channel has many qubit-light transmitters, firing in sequence, while the other side has a small number of qubit-light receivers that collect any arriving signals that overcome loss~\cite{Munro2010}.  The drawback with both of these approaches is that they require a large number of memory qubits or storage modes to increase the number of signals sent through the channel, which may be difficult to engineer.

%\section{Results}
We propose a new communication scheme, called ``midpoint source,'' which uses mature optics technology and is more resilient to loss than midpoint interference.  Our scheme uses two entanglement-swapping operations instead of one.  At the channel midpoint, a triggered source of entangled photons sends each half of an entangled-photon pair in two directions to remotely separated quantum-dot qubits, as shown in Fig.~2.  For conceptual demonstration, we show polarization entanglement here, but Appendix~\ref{App_time_bin} shows how to adapt the scheme to time-bin entanglement, which is preferred for fiber transmission.  The optical channel has two BSM interference apparatuses, each located adjacent to a quantum-dot qubit at either end of the channel.  Each quantum dot is optically excited to produce a photon that is entangled to its spin state at the arrival time of the photon from the channel midpoint.  The two inputs to each BSM are one photon from the entangled-pair source and one photon from the quantum dot.  A successful double-click detection event indicates that entanglement swapping was performed for that side of the channel.  The remotely separated quantum-dot qubits are entangled when entanglement swapping succeeds for both photons from the same entangled pair.  Appendix~\ref{App_SPS} also shows how to replace the entangled-pair source with two single-photon sources, which may have practical advantages.

\begin{figure*}
  \centering
  \includegraphics[width=\textwidth]{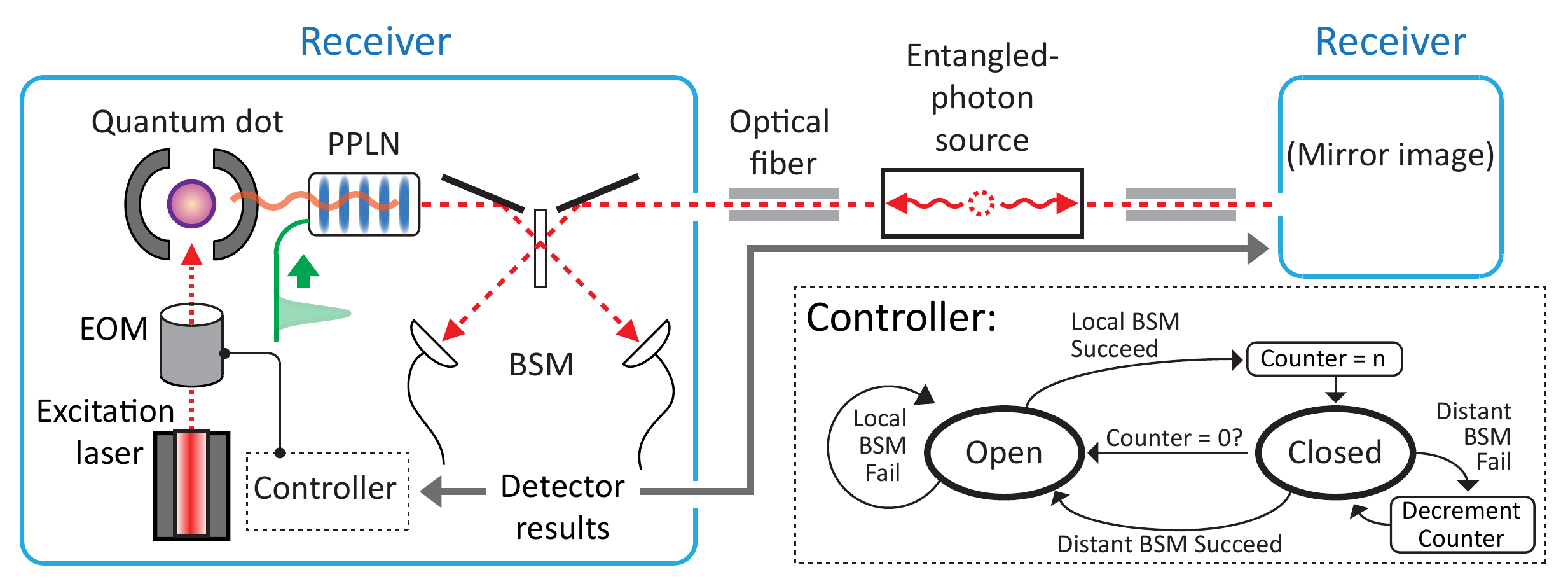}
  \caption{Midpoint-source scheme.  Entangled-photon pairs are generated at the center of the channel and propagate to the receivers.  The station at the endpoint of the channel is a ``receiver,'' which contains many of the elements from Fig.~1.  Each receiver consists collectively of a quantum dot that can produce spin-photon pairs, downconversion of photons to telecom wavelength, and a Bell-state measurement apparatus.  The controller protocol uses information from local and distant BSMs to determine when to excite the quantum dot into a spin-photon entangled state.  Since one of the BSMs is physically adjacent to the controller, a lost photon in the adjacent half of the channel is communicated to the controller with much shorter delay than the scheme in Fig.~1.  The right receiver (not fully depicted) is a mirror image of the left.}
  \label{Midpoint_source}
\end{figure*}

The midpoint-source scheme increases the rate of attempting entanglement by using the adjacent BSM apparatus to rapidly reset the quantum dot when a photon is lost.  We call the joint BSM/quantum-dot apparatus a ``receiver,'' as in Fig.~2.  The receivers use classical signals to synchronize the BSM and communicate BSM results.  The midpoint source generates entangled photons in regular time bins much shorter than $\tau_t$, meaning many photons are sent before the first arrives.  The dynamics of the receiver are governed by the control protocol in Fig.~2, which makes a transition every time step where a photon might arrive.  The states ``open'' and ``closed'' refer to whether the receiver attempts entanglement swapping, which is governed by whether the EOM permits the laser to excite the quantum dot, producing an entangled spin-photon pair.

The control protocol holds the current state in the quantum dot when the local BSM succeeds (indicating swap of entanglement), thereby postselecting events that have overcome half of the optical loss.  Otherwise, the quantum dot resets immediately on local-BSM failure, as in Fig.~3a, which avoids delays inherent to the midpoint-interference scheme.  Two separate spin qubits are entangled only through successful BSM's connected to both halves of the same entangled-photon pair (Fig.~3b).  When at least one BSM fails due to loss, the protocol in Fig.~2 ensures that the receivers return to attempting entanglement as soon as possible.  We presume that successfully entangled spins are used immediately, such as for entanglement swapping in a repeater~\cite{Briegel1998}, to permit resetting the quantum dots.

\begin{figure*}
  \includegraphics[width=\textwidth]{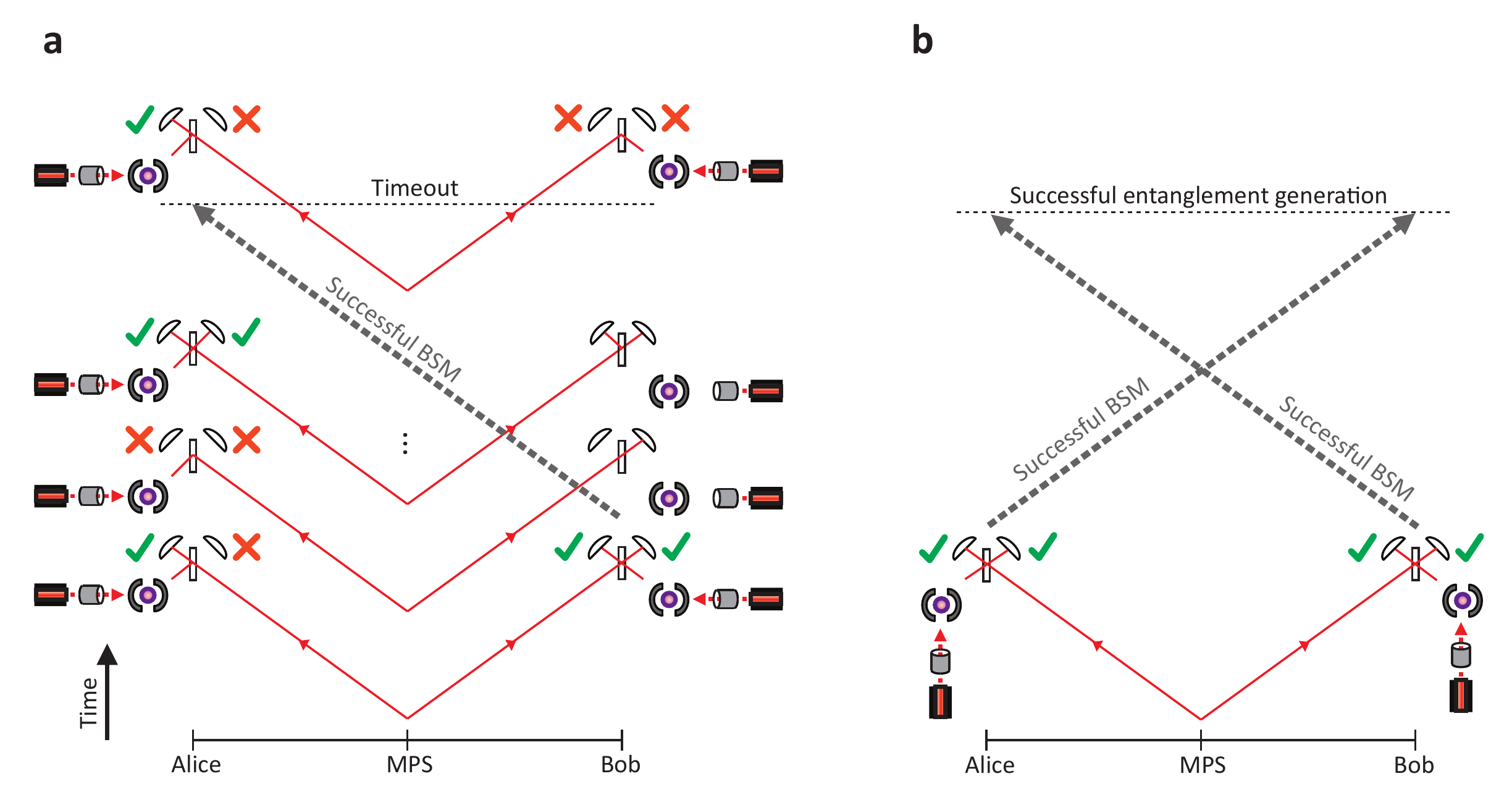}
  \caption{Example timing diagram.  \textbf{a}, Photons emitted by the midpoint source (MPS) travel to receivers at Alice and Bob.  In the first round, the left receiver fails at entanglement swapping, but the right receiver has a successful double-click event.  A classical message is sent from Bob to Alice to announce swapping success.  In subsequent rounds, Alice continues to attempt Bell-state measurements while Bob closes his EOM to preserve the quantum-dot spin.  When Bob's message reaches Alice, she immediately resets her quantum dot; simultaneously, Bob resets due to not receiving a success message from Alice (``timeout'').  \textbf{b}, When both Alice and Bob swap entanglement with entangled photons from the MPS, each announces success and becomes aware of the other side's success after transmission time through the channel.}
  \label{Timing_diagram}
\end{figure*}

At the midpoint source, entangled-photon pairs are generated every clock cycle, $\tau_c \ll \tau_t$.  The clock cycle is limited only by the communication time between BSM and quantum dot, which could in principle be of order 1~ns for a separation of centimeters.  However, we show below that longer cycle times are acceptable.  A receiver is also clocked by $\tau_c$, because it performs entanglement attempts for each arriving photon if the previous BSM failed.  However, the receiver stops attempting BSM's after a success because it must wait time $\tau_t$ to confirm BSM success at the other side of the channel.  For the complete system, the loss in dB is $\alpha_2 \ge \alpha_1$ because of the increased complexity of using two BSM's.  Accordingly, $\beta_2 = 10^{-\alpha_2/10}$.

% If length is a problem, move this paragraph (or at least part of it) to the Supplement
The combination of control protocol and faster clock rate increases throughput.  For simplicity, we assume a symmetric channel in Fig.~2, with equal loss and delay on each side of the midpoint source.  The rate of entanglement generation using our scheme is $G_2 = n \beta_2/[\tau_t(1+n(2\sqrt{\beta_2}-\beta_2))]$, where $n = \tau_t/\tau_c$ (see Methods).    In the limit of $n \rightarrow \infty$, entanglement generation saturates at upper bound ${G_2}^* = \sqrt{\beta_2}/[\tau_t(2-\sqrt{\beta_2})]$.  In regimes of high loss (i.e. $\alpha_2 > 20$ dB), ${G_2}^* \approx \sqrt{\beta_2}/(2\tau_t)$.  Using the control protocol in Fig.~2 with a fast clock cycle, the midpoint-source scheme is only sensitive to loss for half of the channel (specifically, to loss of the other half of the entangled pair that was not postselected by the local BSM measurement---hence the square root).  For the high-loss approximation, the potential improvement factor over midpoint interference is ${G_2}^*/G_1 \approx \sqrt{\beta_2}/(2\beta_1)$, which can be as much as 100 in practice.  Moreover, finite and technologically feasible values of $n$ work sufficiently well.  To achieve $90\%$ of maximum performance, we require $G_2/{G_2}^* = n(2\sqrt{\beta_2}-\beta_2)/[1+n(2\sqrt{\beta_2}-\beta_2)] > 0.9$, or $n(2\sqrt{\beta_2}-\beta_2) > 9$.  For high loss, the condition $n = 5/\sqrt{\beta_2}$ suffices, which approximately corresponds to $\tau_c \approx 500$~ns for $\tau_t = 250$~$\mu$s and $\alpha_2 = 40$~dB.  Hence we may only require 100~ns switching times in electro-optic components, which is achievable using commercial devices.

The performance of the midpoint-source and midpoint-interference schemes are compared as a function of link distance in Fig.~4, using two combinations of loss parameters that are informed by recent experiments~\cite{Gao2012,DeGreve2012,Schaibley2013}.  The communication rate for midpoint source decreases less steeply with distance, because communication rate depends on just half of the channel loss.  The advantage applies to loss from both optical fiber and coupling losses associated with optical quantum dots.  For asymmetric channels, throughput is limited by the side with higher loss.   At distance $L = 50$~km, the communication rate (Bell pairs per second) for our midpoint-source protocol outperforms midpoint interference by a factor of $10-100\times$.  The relative performance advantage increases with communication distance.  Like midpoint interference, the midpoint-source scheme could incorporate multi-mode/multi-qubit storage to further enhance communication rate~\cite{Simon2007,Munro2010}.

\begin{figure}
  \centering
  \includegraphics[width=8.3cm]{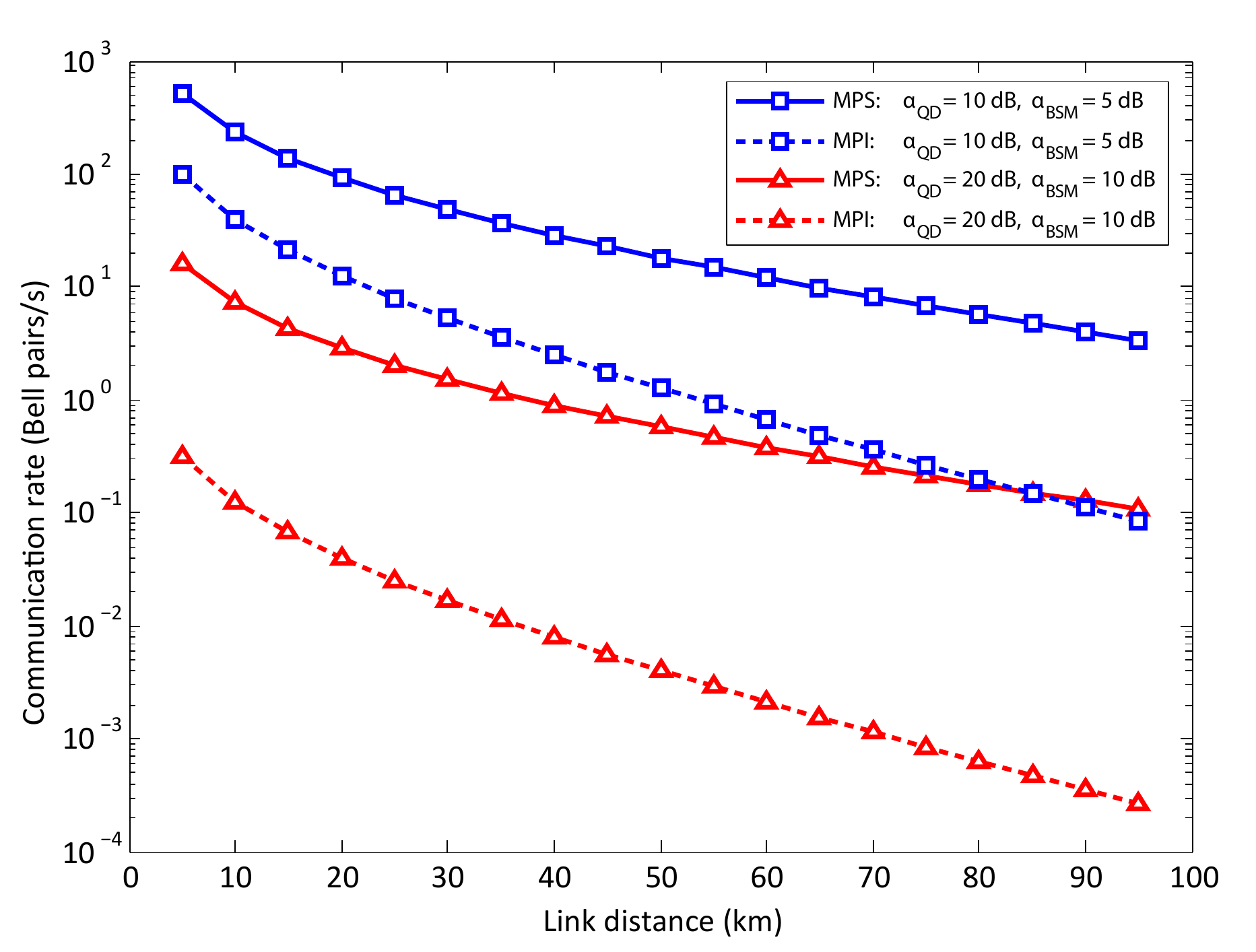}
  \caption{Entangled spin pairs generated per second as a function of distance.  Midpoint-source (MPS; solid lines) and midpoint-interference (MPI; dashed lines) protocols are compared using two sets of fixed optical-loss parameters.  The link distance between quantum dots introduces transmission delay (5~$\mu$s/km) and loss from attenuation in optical fiber (0.2~dB/km).  Square markers correspond to losses from each quantum dot (coupling and frequency conversion) of $\alpha_{\mathrm{QD}} = 10$~dB and BSM losses (partial BSM and imperfect detectors) of $\alpha_{\mathrm{BSM}} = 5$~dB.  Triangle markers correspond to $\alpha_{\mathrm{QD}} = 20$~dB and $\alpha_{\mathrm{BSM}} = 10$~dB.  The midpoint-source scheme delivers communication rates 10-100$\times$ that of midpoint interference, and the slopes of the curves illustrate the square-root sensitivity to loss for midpoint source.  Calculations for midpoint-source performance are explained in Appendix~\ref{App_Markov}.}
  \label{Entanglement_rates}
\end{figure}

%\section{Discussion}
The increased complexity of our entanglement scheme is justified when one considers practical engineering issues that degrade performance of all protocols.  To accommodate photon loss, entanglement is postselected using detection events, which forces the quantum-dot qubits to remain idle until the classical results are available.   Recent experiments~\cite{Moehring2007,Gao2012,DeGreve2012,Schaibley2013,Bernien2013} suggest photon loss might be 40--60~dB, so one must consider protocols tolerant to such high loss.  Our proposed scheme places most of the burden of distributing entanglement on a source of entangled photon pairs.  Recent work has demonstrated high-fidelity sources of 1550-nm entangled photons with entanglement generation rates around $10^6$~pairs/s, so this is a reasonable modification to the system~\cite{Pomarico2012}.  Furthermore, recent improvements to 1550-nm single-photon detectors~\cite{Marsili2013} suggest that the additional photon loss from using a second BSM apparatus does not outweigh the benefits of using a fast clock cycle.

Our proposed scheme is within reach of state-of-the-art experiments and delivers a large enhancement in entanglement generation rates in the presence of high loss and long transmission delay.  This high-speed link would dramatically improve the performance of quantum networks; however, this scheme focuses on just a single entangled link, so some components of a network, like two-qubit gates and long-term memory, are not required for experimental verification.  For quantum dots in particular, our proposal generates higher count rates than midpoint interference, and count rates are already a concern in recent experiments over short distances such as 10~m~\cite{Moehring2007,Gao2012,DeGreve2012,Schaibley2013,Bernien2013}.  For transmission in optical fiber over distances appropriate for quantum networks, our proposal improves communication rates by about $10$--$100\times$ over that of previous schemes.  Finally, the resilience of the midpoint-source scheme to loss and delay makes it ideal for distributing entanglement through the atmosphere using an entangled-photon source on a satellite~\cite{Aspelmeyer2003}.

\begin{methods_summary}
The proposed implementation with quantum dots is based on recent experiments demonstrating the generation of photons entangled to a quantum-dot spin~\cite{Gao2012,DeGreve2012,Schaibley2013}.  Although InAs quantum dots generate photons with wavelength around 900~nm, the photons can undergo coherent frequency conversion~\cite{Ates2012,Zaske2012,DeGreve2012} to telecom-band wavelength (e.g. 1550~nm), for which attenuation in optical fiber is minimized at approximately 0.2~dB/km.  Even if the BSM is performed using free space optics, the converted quantum-dot photons must be tuned such that they are indistinguishable from photons generated by the entangled-pair source~\cite{Pomarico2012}, which will be at a telecom wavelength.  Additionally, the frequency conversion process can potentially compensate for quantum dots having different emission wavelengths.

The control protocol for the midpoint-source scheme (Fig.~2) is a finite-state machine that evolves in discrete time steps as a Markov chain.  A state transition occurs every clock cycle $\tau_c$.  Each transition corresponds to a detection event, and the associated probability for each outcome is determined by optical loss and detector efficiency.  The expected rate of generating entanglement can be calculated from the equilibrium distribution on this Markov chain, as detailed in Appendix~\ref{App_Markov}.
\end{methods_summary}

\begin{acknowledgments}
This work was supported by the JSPS through its FIRST Program, NICT, and Special Coordination Funds for Promoting Science and Technology. CJ acknowledges support as an NSF Graduate Fellow. KDG acknowledges support as a Herb and Jane Dwight Stanford Graduate Fellow and as a Harvard Quantum Optics Center (HQOC) postdoctoral fellow.
\end{acknowledgments}

\appendix
\section{Bell-state measurement apparatus}
\label{App_BSM}
\begin{figure}
  \includegraphics[width=8cm]{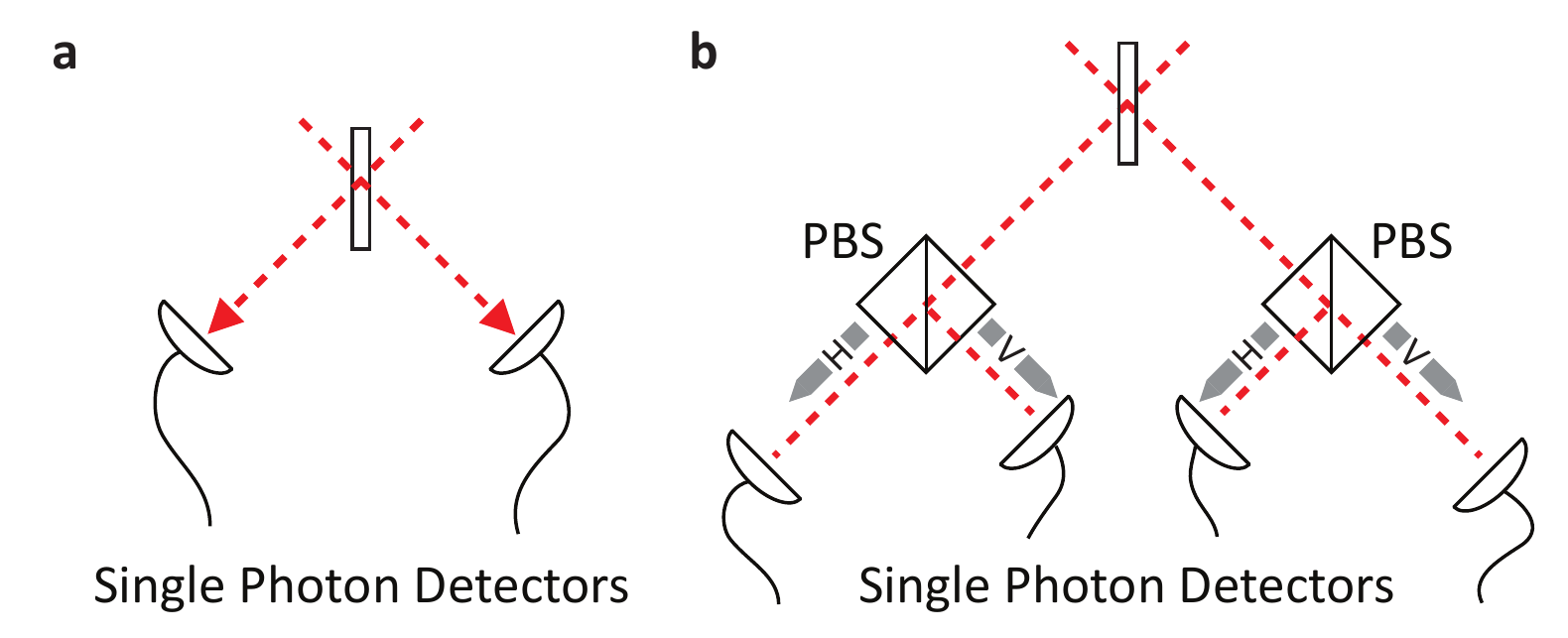}
  \caption{Bell-state measurement (BSM) apparatuses.  \textbf{a},~Two-detector arrangement that can identify the singlet state.  \textbf{b},~Four-detector arrangement that can identify singlet and one triplet state.}
  \label{fig::BSM_apparatus}
\end{figure}

In the simplest case, the Bell-state measurement (BSM) apparatus consists of a beam splitter with two input ports and two output ports which feed into single-photon detectors, as in Fig.~\ref{fig::BSM_apparatus}a below.  If two indistinguishable photons arrive simultaneously on different input ports, they bunch and exit at the same port; only one of the detectors ``clicks.''  We are interested in the case where two distinguishable photons arrive.  Let us suppose the photons are distinguishable only by polarization degrees of freedom, with a basis spanned by horizontally polarized $\ket{H}$ and vertically polarized $\ket{V}$.

When distinguishable single photons arrive at the BSM from both sides, they are in some superposition of the triplet state $2^{-1/2}(\ket{H_L V_R} + \ket{V_L H_R})$ and singlet state $2^{-1/2}(\ket{H_L V_R} - \ket{V_L H_R})$, where subscripts denote left and right input ports.  If the photons are in the triplet state, then they bunch and exit the same port.  If instead the photons are in the singlet state, then they exit different output ports.  In this way, a ``double-click'' event (two detectors fire simultaneously) performs projective measurement into the singlet state, which can be used for entanglement swapping~\cite{Zukowski1993,Briegel1998,Pan1998,Duan2001,Moehring2007,Chen2007}.  One can also measure the triplet state $2^{-1/2}(\ket{H_L V_R} + \ket{V_L H_R})$ by placing a polarizing beam splitter at each output port and using four detectors~\cite{Michler1996} (Fig.~\ref{fig::BSM_apparatus}b).  The ``double click'' pattern in this case is any pair of detectors aligned with output ports for different polarizations.  Whether the two detectors are on different sides or the same side of the non-polarizing beam splitter distinguishes singlet and triplet states, respectively.

%Crucially, a photon emitted by the ``left'' quantum dot in state $\ket{H}$ must be indistinguishable from a photon in state $\ket{H}$ emitted by the ``right'' quantum dot, and likewise for photons in state $\ket{V}$.  If this condition does not hold (such as if the photons in the same polarization have different frequencies due to quantum-dot inhomogeneity), then the beam splitter does not erase ``which-path'' information for the photon.  If, after passing through the beam splitter, a photon could be traced back to a specific quantum dot, then the BSM fails since leakage of which-path information to the environment introduces decoherence.

For the remaining states where indistinguishable photons arrive ($\ket{H_L H_R}$ or $\ket{V_L V_R}$), the photons bunch at the output of the beam splitter and trigger only one detector.  These two states cannot be discerned from each other using this apparatus.  By post-selecting events which generate a ``double click,'' the BSM implements probabilistic entanglement swapping with identified failure events.  Without impacting the entanglement-swapping procedure, the failure events are counted as loss, since they are useless for entanglement distribution.  The BSM apparatus can only detect two of the four Bell states with linear optics, achieving the maximum success probability 50\% with ideal detectors~\cite{Luetkenhaus1999,Calsa2001}.  For this reason, the apparatus is sometimes called a ``partial BSM.''  Imperfect detectors with quantum efficiency $\eta < 1$ will introduce more loss, but recent advances in single-photon detectors suggest that $\eta$ around 0.3 to 0.5 is plausible~\cite{Natarajan2012,Marsili2013}.

Besides finite quantum efficiency, realistic experiments will have to contend with ``dark counts''---false-positive events where the detector fires despite no photon arriving.  If one of the photons is lost, but a double click occurs because of a dark count, then entanglement swapping fails, and the distant spins are in a mixed state.  The possibility that any BSM double click could be due to a dark count degrades entangled-state fidelity, so dark count rates must be suppressed as much as possible.  Superconducting nanowire single photon detectors can achieve dark counts below 100 per second and timing jitter less than 100~ps~\cite{Natarajan2012}.  Since the temporal length of photons from InAs quantum dots is around 1~ns~\cite{Gao2012,DeGreve2012,Schaibley2013}, one could filter the signal output of the detector in, for example, 10~ns windows.  The effective dark-count probability would be $p_{\mathrm{dc}} = (10 \;\mathrm{ns})(100 \;\mathrm{s}^{-1}) = 10^{-6}$ per filtering window.

The probability of dark count per window must be compared with the probability that a photon is lost in one of the optical paths in this same window.  For the midpoint-source scheme, assume both receivers are in the ``open'' state (attempting spin entanglement).  Assuming symmetric loss in the system, denote $\beta_{\mathrm{qd}}$ as the probability for a photon emitted by a quantum dot to be detected, including losses in optical coupling, frequency conversion, and finite detector efficiency.  Similarly, denote $\beta_{\mathrm{ms}}$ as the probability for a photon emitted by the midpoint source to be detected, again accounting for coupling loss, detector inefficiency, and possibly entangled-pair generation probability.  Clearly, the total loss parameter of the main text is given by $\beta_2 = (\beta_{\mathrm{qd}}\beta_{\mathrm{ms}})^2$.

The probability of incorrectly accepting a prospective entangled state due to a dark count is $2 p_{\mathrm{dc}} ({\beta_{\mathrm{qd}}}^2 \beta_{\mathrm{ms}}(1-\beta_{\mathrm{ms}}) + {\beta_{\mathrm{ms}}}^2 \beta_{\mathrm{qd}}(1-\beta_{\mathrm{qd}}))$, to lowest order in $p_{\mathrm{dc}}$.  If $p_{\mathrm{dc}} \ll \mathrm{min}\{\beta_{\mathrm{qd}},\beta_{\mathrm{ms}}\}$, then the higher-order terms are negligible.  The preceding expression applies to the partial BSM that detects only the singlet state; for the BSM that detects both singlet and one of the triplets, there is an additional factor of 2 to account for false-positive outcomes resulting from two possible detectors that could have a dark count.  In either case, the fidelity of the spin-spin entangled state heralded by double-click events in both receivers is approximately given by
\begin{eqnarray}
1-F & \sim & \frac{p_{\mathrm{dc}} ({\beta_{\mathrm{qd}}}^2 \beta_{\mathrm{ms}}(1-\beta_{\mathrm{ms}}) + {\beta_{\mathrm{ms}}}^2 \beta_{\mathrm{qd}}(1-\beta_{\mathrm{qd}}))}{(\beta_{\mathrm{qd}}\beta_{\mathrm{ms}})^2} \nonumber \\
& = & \frac{p_{\mathrm{dc}}(\beta_{\mathrm{qd}}(1-\beta_{\mathrm{ms}})+\beta_{\mathrm{qd}}(1-\beta_{\mathrm{ms}}))}{\beta_{\mathrm{qd}}\beta_{\mathrm{ms}}}.
\end{eqnarray}
If $p_{\mathrm{dc}} \ll \mathrm{min}\{\beta_{\mathrm{qd}},\beta_{\mathrm{ms}}\}$, then the fidelity is close to 1.  This condition is satisfied by experimentally realistic values like $\beta_{\mathrm{qd}} \sim 10^{-2}$ and $\beta_{\mathrm{ms}} \sim 10^{-2}$.

As a side note, the foregoing analysis suggests that the midpoint-source scheme could be more robust to dark counts than the midpoint-interference scheme.  For the latter, the loss for each half of the channel is $\sqrt{\beta_1}$, which is larger than $\beta_{\mathrm{qd}}$ because the two schemes have the same sources of loss for these terms, except the midpoint-interference scheme must also include attenuation in optical fiber.  As a result, the fidelity of midpoint-interference, which is approximately given by
\begin{equation}
1-F \sim \frac{p_{\mathrm{dc}} \sqrt{\beta_1}}{\beta_1},
\end{equation}
can be more sensitive to dark counts.  The intuitive explanation is that the midpoint-source scheme divides lossy operations over four detection events, whereas midpoint-interference combines lossy paths into just two detection events.  As a result, the photon flux going into each detector for the midpoint-source scheme is brighter, diminishing the impact of dark counts.  This observation applies even with the use of gating techniques to suppress dark count rates.  The advantage disappears if the midpoint-source scheme has a high-loss component not also in the midpoint-interference scheme, such as an entangled-photon source with very low pair-generation efficiency.

\section{Photons with time-bin entanglement}
\label{App_time_bin}
Time-bin entanglement could be more robust for fiber transmission than polarization~\cite{Honjo2007,Dynes2009}.  The working principle is the same as outlined above for polarization entanglement.  The single photons emitted by both quantum dots into the ``early'' time bin must be indistinguishable, and likewise for the photons emitted into the ``late'' time bin.  Again, photons pass through a beam splitter into detectors.  In this configuration, the detectors require sufficient timing resolution to distinguish ``early'' and ``late'' photon states.  A ``double-click'' event for the singlet state is two different detectors signal at different times.  If a detector has sufficiently fast recovery time, then the triplet state can be detected by two separate detection signals on the same detector, meaning the photons exit the same port.  As with polarization-encoded photons, the maximum success probability is 50\% for entanglement swapping.

\begin{figure}
  \includegraphics[width=6cm]{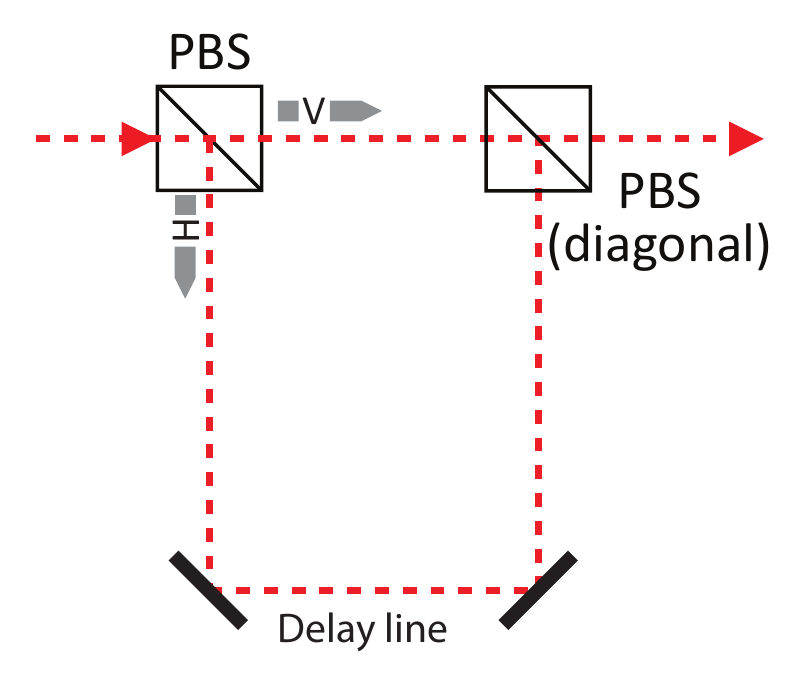}
  \caption{Apparatus for converting a photonic qubit with polarization degree of freedom to time-bin degree of freedom.  The first polarizing beam splitter (PBS) separates the two possible states of the photon into different rails, where one rail has a delay from a longer path.  The two rails are recombined at another PBS placed at a diagonal orientation so as to erase the polarization degree of freedom.  This procedure introduces a loss of 50\% since half the photons will exit the top port instead of the right port.}
  \label{fig::time_bin}
\end{figure}

There are simple ways to transform polarization entanglement to time-bin entanglement.  Using a polarizing beam splitter (PBS), the polarization states are split into two rails.  One rail has a longer optical path that introduces a delay, as in Fig.~\ref{fig::time_bin}.  The two rails are recombined using another PBS at diagonal orientation.  Using only light that exits one of the PBS ports, the polarization information is erased, but half of the photons are lost.  Whether the advantages of time-bin entanglement offset the additional loss depends on implementation details not considered here.

\section{Single-photon sources at midpoint}
\label{App_SPS}
\begin{figure}
  \includegraphics[width=6cm]{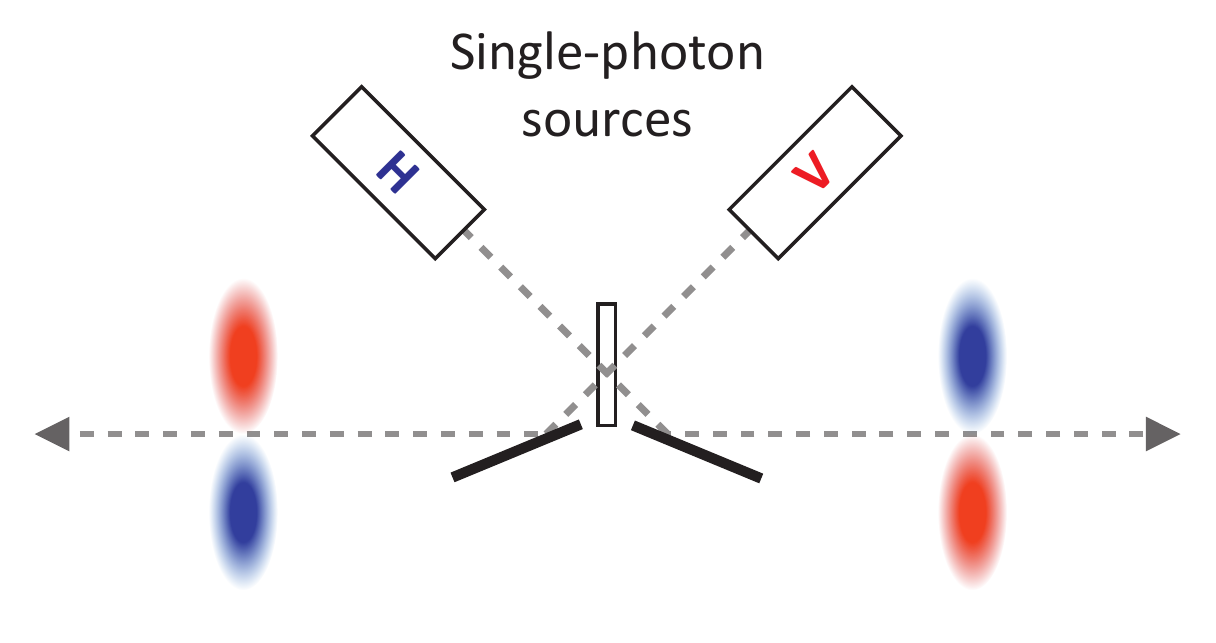}
  \caption{Apparatus for replacing entangled-photon source with single-photon sources and a beam splitter.  Polarization photon states are shown for concept, but the method also applies to time-bin entanglement.  Double-click events on both sides of the channel imply that one photon went in each direction.  Without knowledge of which photon went in each direction (illustrated here with possibilities above and below the dashed line), the entangled singlet state is post-selected.}
  \label{fig::SPS}
\end{figure}

The midpoint source of entangled-photon pairs can be replaced by two single-photon sources~\cite{Fattal2004}.  As explained below, this strategy suffers an additional factor of 50\% loss in each transmission compared with the entangled-photon source, but this penalty may be offset if single-photon sources prove easier to engineer at high clock rates.  Let us work in the polarization basis of a photonic qubit with states horizontal (H) and vertical (V).  The setup in Fig.~\ref{fig::SPS} has two identical single-photon sources, except one is oriented at horizontal polarization and the other vertical, and each source is coupled into one of the two input ports of a beam splitter.  The two sources are triggered simultaneously to emit a photon.  Because the two photons are distinguishable, there is no bunching.  The state after the beam splitter is $2^{-1}\left(\ket{H_L V_L} + \ket{H_L V_R} - \ket{V_L H_R} - \ket{H_R V_R}\right)$, where subscripts denote left and right output ports.  If the Bell-state measurement apparatuses on both sides of the channel register ``double clicks'', then it must be the case that one photon went in each direction from the midpoint, assuming ideal detectors and no light pollution.  This post-selects the entangled-photon state $2^{-1/2}\left(\ket{H_L V_R} - \ket{V_L H_R} \right)$, which facilitates the entanglement of remotely separated quantum dots.  Interestingly, the photons were not entangled in flight, though entanglement swapping is produced through projective measurement.  A similar proposal for the use of sub-Poissonian light was made in the context of quantum key distribution~\cite{Fattal2004}.  Because the entangled-photon state is post-selected from a separable state, the success probability of entangling the two quantum dots is reduced by a factor of 50\%, which is precisely the probability overlap of the separable and singlet photon states.  This scheme can be adapted to time-bin entanglement using just one single-photon source.  The source emits two photons in sequence into a beam splitter whose output ports couple into each direction of the channel.  Successful time-bin double clicks on both sides will post-select the time-bin singlet state.

\section{Control protocol and average rate of entanglement}
\label{App_Markov}
The time-dependent behavior of the control protocol for our entanglement-distribution scheme can be modeled using a Markov chain.  We need to determine the average rate at which entanglement succeeds as a function of channel loss ($\beta_2$), channel transmission delay ($\tau_t$), and clock cycle time ($\tau_c$).  As before, let $n = \tau_t/\tau_c$ be an integer, although $n = \lceil \tau_t / \tau_c \rceil$ would suffice in general.  Also, we assume the channel is symmetric in loss and delay.  This simplifies analysis, but the protocol can be adapted to asymmetric parameters.

\begin{figure}
  \centering
  \includegraphics[width=8.3cm]{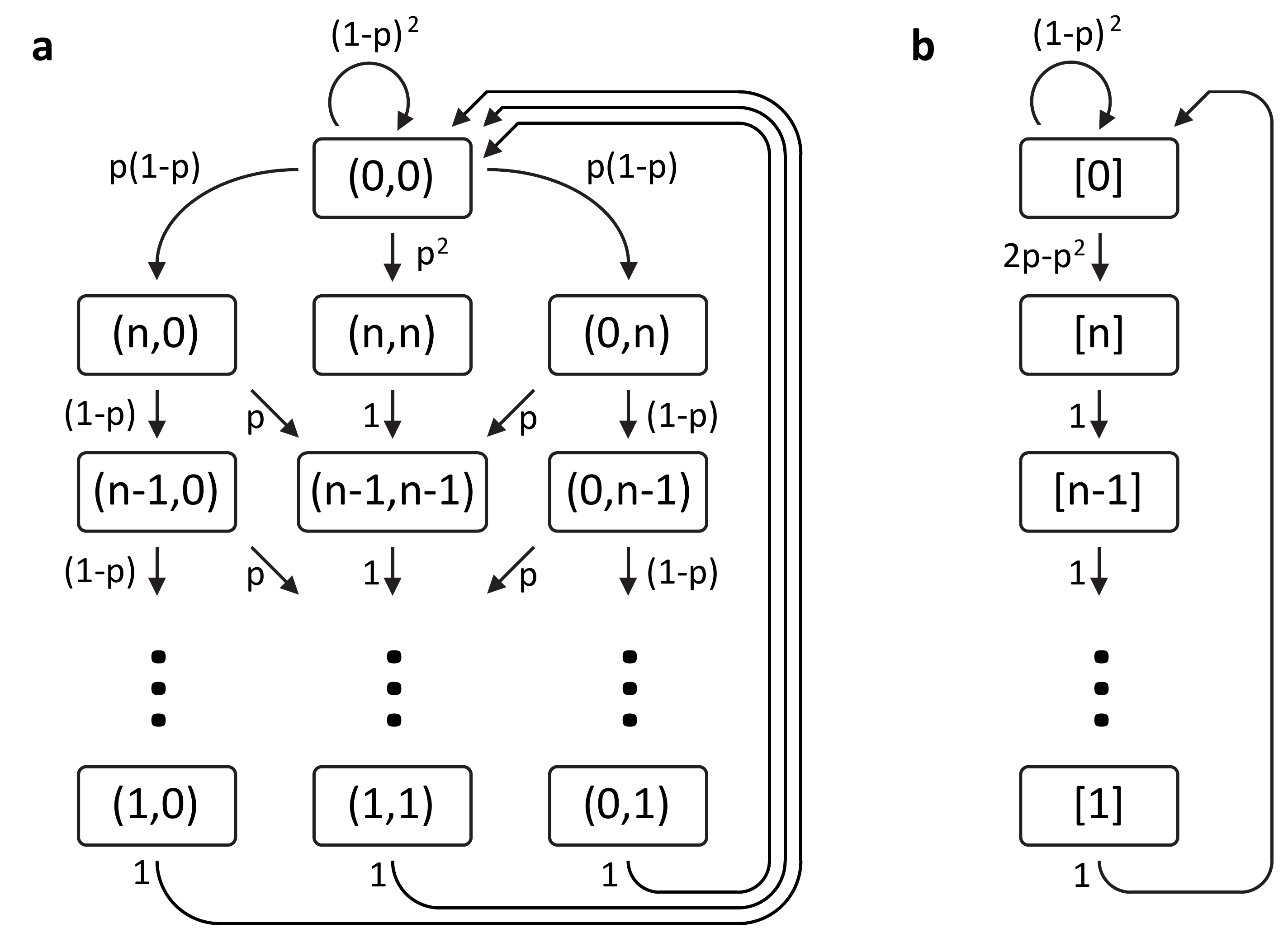}
  \caption{Markov chains representing the time-dependent behavior of the entanglement protocol.  \textbf{a},~A Markov chain where each state $(x,y)$ denotes the number of remaining cycles that left and right receivers will be closed, respectively, based on BSM detection events.  \textbf{b},~A simplified Markov chain where each state $[x]$ denotes the time until both receivers are open.  The equilibrium distributions of both chains have the same probability of occupying the $(0,0)$ and $[0]$ states, but the latter is simpler to calculate.}
  \label{Markov_chains}
\end{figure}

The Markov chain has a discrete state space and makes a transition every clock cycle.  There are $3n+1$ states which are labeled as: $(0,0)$, both receivers are ``open''; $\{(i,0)\}_{i=1}^n$, the left receiver had a successful Bell-state measurement (BSM) and is ``closed'' for $i$ more cycles; $\{(0,i)\}_{i=1}^n$, the right receiver had a successful BSM and is ``closed'' for $i$ more cycles; $\{(i,i)\}_{i=1}^n$, both receivers had a successful BSM (though possibly not in the same cycle), and they are both ``closed'' for $i$ more cycles.  Transitions between states are dictated by the control protocol and photon detection probabilities.  The Markov chain associated with this behavior is shown in Fig.~\ref{Markov_chains}a, where we use $p = \sqrt{\beta_2}$, the probability of a successful BSM outcome in each half of a channel with symmetric loss.

The state transition probabilities in Fig.~\ref{Markov_chains}a are determined by the control protocol given in Fig.~2 of the main text.  In particular, if only one side of the channel has successful BSM, the receiver will ``time out'' after $n$ cycles.  If one side of the channel has successful BSM, then subsequently receives a success signal from the other side, but associated with a different time bin, then this receiver immediately resets.  At the same time, the other side of the channel will automatically time out, because it has waited for $\tau_t$, the duration of transmission delay.  The Markov chain captures this behavior by assuming error-free classical communication that is limited by propagation of signals in optical fiber at approximately $2\times10^8$ m/s.  Specifically, the Markov chain acts as an omniscient observer that knows when either side of the channel will reset, even though the appropriate messages have not yet arrived.

The average rate of entanglement generation can be determined using the equilibrium distribution on the Markov chain.  We denote the steady-state probability of being in state $(0,0)$ as $\pi_{(0,0)}$, etc.  The only parameter we are interested in is $p^2 \pi_{(0,0)}$, the probability flux from $(0,0)$ to $(n,n)$, which corresponds to the event of successful, coincident BSM's at both receivers.  This quantity is the average number of successful entanglement events per clock cycle, so $G_2 = \beta_2 \pi_{(0,0)}/\tau_c$ is the entanglement rate in real units.

To calculate the equilibrium distribution, we introduce a simplified Markov chain.  Denote simplified state space $\{[i]\}_{i=0}^n$ where occupation probability $s_{[i]}$ is related to the original Markov chain by $s_{[i]} = s_{(i,0)} + s_{(0,i)} + s_{(i,i)}$, except $s_{[0]} = s_{(0,0)}$.  Simply put, each row of states in Fig.~\ref{Markov_chains}a has been collapsed into a single state, as in Fig.~\ref{Markov_chains}b.  In the simplified Markov chain, $i>0$ represents the number of ``closed'' cycles until both receivers are ``open'' again.  The equilibrium conditions for the simplified Markov chain are linearly dependent on the equilibrium conditions for the first chain, so the steady-state probability in the $[0]$ state is $\pi_{[0]} = \pi_{(0,0)}$.  The simplified Markov chain is finite, irreducible, and aperiodic, so a unique equilibrium distribution exists.  Moreover, by simple application of equilibrium conditions, we can show that $\pi_{[i]} = (1-\pi_{[0]})/n$, for all $i > 0$.  Finally, $\pi_{[0]} = 1/(1 + n (2p-p^2))$.  In the main text, we express entanglement rate as $G_2 = n \beta_2/[\tau_t(1+n(2\sqrt{\beta_2}-\beta_2))]$.

%\bibliography{Entanglement_references}
%

\end{document}